\documentclass[fleqn,12pt]{wlscirep}
\usepackage{diagbox}
\usepackage[symbol]{footmisc}

\usepackage{graphicx} 
\graphicspath{{figures/}} 

\newcommand{\mgii}{Mg\ensuremath{\,\textsc{ii}}}
\newcommand{\cii}{[C\ensuremath{\,\textsc{ii}}]}

\newcommand{\civ}{C\ensuremath{\,\textsc{iv}}}

\newcommand{\farcs}{\mbox{$.\!\!^{\prime\prime}$}}%

 \newcommand{\kms}{{\rm km\,s}\ensuremath{^{-1}}}

\newcommand{\Jntt}{\ensuremath{J_{\rm SOFI}}}
\newcommand{\Hntt}{\ensuremath{H_{\rm SOFI}}}
\newcommand{\Kntt}{\ensuremath{Ks_{\rm SOFI}}}

\newcommand{\blazar}{J0410--0139}

\newcommand\aj{Astron. J.}
\newcommand\araa{Annu. Rev. Astron. Astrophys.}
\newcommand{\aap}{Astron. Astrophys.}               
\newcommand\mnras{Mon. Not. R. Astron. Soc.}%
\newcommand\apjs{Astrophys. J. Suppl. Ser.}%
\newcommand\apj{Astrophys. J}%
\newcommand\apjl{Astrophys. J}%
\newcommand\pasp{Publ. Astron. Soc. Pac}%
\newcommand{\pasa}{Public. of the Astron. Soc. of Australia}

\title{A blazar in the epoch of reionization}

\author[1,*]{Eduardo Ba\~nados}
\author[2]{Emmanuel Momjian}
\author[3]{Thomas Connor}

\author[1,4]{Silvia Belladitta}
\author[4]{Roberto Decarli}
\author[5]{Chiara Mazzucchelli}
\author[6]{Bram P.\ Venemans}
\author[1]{Fabian Walter}
\author[7]{Feige Wang}
\author[1]{Zhang-Liang Xie}

\author[8]{Aaron J.\ Barth}
\author[9]{Anna-Christina Eilers}
\author[7]{Xiaohui Fan}
\author[1]{Yana Khusanova}
\author[10]{Jan-Torge Schindler}
\author[11]{Daniel Stern}
\author[7]{Jinyi Yang}

\author[12,13]{Irham Taufik Andika}
\author[2]{Christopher L.\ Carilli}
\author[14]{Emanuele P.\ Farina}
\author[15]{Andrew Fabian}
\author[6,16]{Joseph F.\ Hennawi}
\author[17]{Antonio Pensabene}
\author[1,18]{Sof\'ia Rojas-Ruiz}

\affil[1]{Max-Planck-Institut f\"ur Astronomie, K\"onigstuhl 17, D-69117, Heidelberg, Germany}
\affil[2]{National Radio Astronomy Observatory, Pete V. Domenici Science Operations Center, P.O. Box O, Socorro, NM 87801, USA}
\affil[3]{Center for Astrophysics $\vert$\ Harvard\ \&\ Smithsonian, 60 Garden Street, Cambridge, MA 02138, USA}
\affil[4]{INAF -- Osservatorio di Astrofisica e Scienza dello Spazio, via Gobetti 93/3, 40129, Bologna, Italy}
\affil[5]{Instituto de Estudios Astrof\'{\i}sicos, Facultad de Ingenier\'{\i}a y Ciencias, Universidad Diego Portales, Avenida Ejercito Libertador 441, Santiago, Chile}
\affil[6]{Leiden Observatory, Leiden University, Niels Bohrweg 2, 2333 CA Leiden, Netherlands}
\affil[7]{Steward Observatory, The University of Arizona, 933 North Cherry Avenue, Tucson, AZ 85721--0065, USA}
\affil[8]{Department of Physics and Astronomy, 4129 Frederick Reines Hall, University of California, Irvine, CA, 92697-4575, USA}
\affil[9]{MIT Kavli Institute for Astrophysics and Space Research, 77 Massachusetts Ave., Cambridge, MA 02139, USA}
\affil[10]{Hamburger Sternwarte, Universität Hamburg, Gojenbergsweg 112, D-21029 Hamburg, Germany}
\affil[11]{Jet Propulsion Laboratory, California Institute of Technology, 4800 Oak Grove Drive, Pasadena, CA 91109, USA}
\affil[12]{Technical University of Munich, TUM School of Natural Sciences, Department of Physics, James-Franck-Str. 1, D-85748 Garching, Germany}
\affil[13]{Max-Planck-Institut f\"{u}r Astrophysik, Karl-Schwarzschild-Str. 1, D-85748 Garching, Germany}
\affil[14]{Gemini Observatory, NSF’s NOIRLab, 670 N A’ohoku Place, Hilo, Hawai'i 96720, USA}
\affil[15]{Institute of Astronomy, Madingley Road, Cambridge CB3 0HA, UK}
\affil[16]{Department of Physics, Broida Hall, University of California, Santa Barbara, CA 93106--9530, USA}
\affil[17]{Dipartimento di Fisica “G. Occhialini”, Università degli Studi di Milano-Bicocca, Piazza della Scienza 3, I-20126, Milano, Italy}
\affil[18]{Department of Physics and Astronomy, University of California, Los Angeles, 430 Portola Plaza, Los Angeles, CA 90095, USA}

\affil[*]{banados@mpia.de}

\begin{abstract}
\textbf{
Relativistic jets are thought to play a crucial role in the formation and evolution of massive galaxies and supermassive black holes. Blazars—quasars with jets aligned along our line of sight—provide insights into the jetted population and have been observed up to redshifts of $z=6.1$. Here, we report the discovery and multi-wavelength characterization of the blazar  VLASS~J041009.05$-$013919.88 at $z=7$ (Universe's age $\sim$750\,Myr), powered by a $\sim 7\times10^8\,M_\odot$ black hole. 
 The presence of this high-redshift blazar implies a large population of similar but unaligned jetted sources in the early Universe. 
Our findings suggest two possible scenarios: in one, the jet in J0410--0139 is intrinsically low-power but appears highly luminous due to relativistic beaming, suggesting that most UV-bright quasars at this redshift host jets. Alternatively, if J0410--0139 represents an intrinsically powerful radio source, there should be hundreds to thousands of radio-quiet quasars at $z\sim7$ with properties similar to J0410--0139—a prediction in tension with observed quasar densities based on their UV luminosity function. These results support the hypothesis that rapid black hole growth in the early Universe may be driven by jet-enhanced or obscured super-Eddington accretion, potentially playing a key role in forming massive black holes during the epoch of reionization.
}
\end{abstract}

\begin{document}

\flushbottom
\maketitle
\thispagestyle{empty} 

Relativistic jets are believed to be key drivers in the growth of massive galaxies and supermassive black holes \cite{blandford2019}, and a few of them have been identified in the first billion years of the Universe \cite{belladitta2020,banados2021}. 
We selected the quasar VLASS J041009.05$-$013919.88 (hereafter \blazar) from crossmatching the optical DESI Legacy Imaging Surveys\cite{dey2019} (DELS) and the 1.4 GHz NRAO VLA Sky Survey\cite{condon1998} (NVSS). We required that the source be undetected in the DELS $g$- and $r$-bands (Fig.~\ref{fig:discovery} and Table~\ref{tab:obs}). The NVSS beam is 45$^{\prime\prime}$; thus, the radio emission can arise from several sources. Therefore, we required that the DELS source coincides with an unresolved radio source in the 3 GHz Very Large Array Sky Survey\cite{lacy2020} (VLASS). The VLASS 2\farcs5 beam unambiguously identified the NVSS radio source with the DELS source (i.e., no other radio sources within the NVSS beam). The 3\,GHz flux density of \blazar\ increased by $\sim35\%$ from the first to the second VLASS epoch (in less than three years observed frame; see Fig.~\ref{fig:discovery} and Table~\ref{tab:radio}); this was the first indication that the source might be variable and potentially a blazar. 

We confirmed the nature of \blazar\ on 7 November 2020 with a 30 min FORS2 spectrum at the Very Large Telescope (VLT) in Chile, revealing a quasar with a prominent Ly$\alpha$ emission line at $z=7.0$---within the so-called epoch of reionization, when the intergalactic medium is transitioning from neutral to ionized\cite{davies2018b,Wang2020ApJ...896...23W}. We then obtained near-infrared spectra with Keck/NIRES, LBT/LUCI, and Magellan/FIRE (see Fig.\ \ref{fig:discovery}; and Methods for details on the observations). The spectra were taken between 2020 November and 2021 November, and the spectral slopes and broad emission line profiles are consistent among the different spectra (Fig.~\ref{fig:discovery}). Near-infrared follow-up photometry taken between 2020 November and 2023 January are also consistent {within $2\sigma$ (Table~\ref{tab:obs}). Thus, we find no evidence for variability in the observed near-infrared wavelengths (rest-frame UV/optical), implying no strong changes in the quasar's accretion disk and broad line region. These characteristics are seen in luminous flat spectrum radio quasars (FSRQ) at lower redshifts\cite{ghisellini2015c}, and consistent with the flat observed 2.7--5\,GHz radio spectrum\cite{urry1995} of \blazar\ (Fig.~\ref{fig:sed}).

The redshift derived from a Gaussian fit to the \mgii\ line is $z_{\rm \mgii} = 6.995\pm 0.001$ (all error bars in this article correspond to $1\sigma$ or the central $68\%$ interval of the distribution, unless otherwise stated), but hereafter we use  $z_{\rm \cii} = 6.9964 \pm 0.0005$ as the systemic redshift, measured from ALMA observations of the \cii\ line presented in a companion paper\cite{banados2024b}. Throughout this paper, we adopt a flat cosmology with a Hubble parameter of $H_0 = 70\, \kms\,\mathrm{Mpc}^{-1}$, and cosmological density parameters $\Omega_{\rm M}=0.3$ and $\Omega_{\Lambda}=0.7$.  In this cosmology, the universe's age at $z_{\rm \cii}$ is 751\,Myr and 1$^{\prime\prime}$ corresponds to $5.23$ proper kpc.

We derive a black hole mass of $6.9^{+0.5}_{-0.4}\times 10^8$ the mass of the Sun ($M_\odot$) from the full-width at half maximum (FWHM) of the \mgii\ line and the quasar luminosity at 3000\,\AA\ ($L_{3000}$; see Table~\ref{tab:properties}), following the methods of previous works\cite{vestergaard2009,fan2023}. 
Adopting a bolometric correction\cite{richards2006b} of $L_{\rm bol} = 5.15\times L_{3000}$, the accretion rate or Eddington ratio for \blazar\ is $L_{\rm bol}/L_{\rm Edd}=1.22 \pm 0.08$. These properties are comparable to what is observed in other quasars at a similar cosmic time but without evidence of powerful relativistic jets\cite{fan2023}.

Observational signatures of a blazar or a relativistic jet pointing close to our line of sight include\cite{ighina2019}: (a) strong variable radio emission; (b) flat or peaked core-dominated radio spectrum; (c) compact/core-jet radio morphology; (d) flat X-ray photon index ($\Gamma_X<1.8)$; (e) flat rest-frame 10\,keV to 2500\,\AA\ flux ratio ($\tilde{\alpha}_{\rm OX}$$>-1.36$). Below we show that \blazar\ satisfies all these criteria.

Fig.~\ref{fig:sed}\textit{a} shows the full spectral energy distribution (SED) of \blazar\ and of low redshift blazars of similar radio and X-ray luminosities\cite{massaro2015}. We compiled all existing archival radio observations of \blazar\ (Table~\ref{tab:obs}), and obtained quasi-simultaneous VLA observations on 25 September 2021 and on 6 August 2022 from 1 to 12\,GHz, corresponding to rest-frame 8 to 96\,GHz (see Methods). The measured flux densities differed from all archival observations, confirming a highly variable radio SED as shown in Fig.~\ref{fig:sed}\textit{b}.  The quasi-simultaneous 2021 VLA observations reveal a turnover at rest-frame $\sim$40\,GHz, making \blazar\ a gigahertz peaked radio source, similar to other blazars\cite{coppejans2017} and indicative of a young jet or a blazar flaring event\cite{tinti2005,odea2021}. The 2022 radio SED shows a flatter spectrum with a less pronounced peak. 
{Fig.~\ref{fig:sed}\textit{c} shows the lightcurves at observed frequencies of 1.4 and 3.0\,GHz, spanning 3.4 years and 8 months in the quasar's rest frame, respectively.
They are incompatible with the expectations of a young radio source in adiabatic expansion\cite{orienti2020,nyland2020}. The flux density exhibited a change by a factor of three in 14 days in the quasar's rest frame. Such variability and change of radio spectral shape are similar or more extreme than that of blazars at lower redshifts\cite{mufakharov2021,komossa2023}. 

On 17 November 2021, we obtained $14.3 \times 5.9$ mas ($ 75 \times 37$ pc) resolution Very Long Baseline Array (VLBA) observations at 1.5\,GHz (see Methods and Fig.~\ref{fig:vlba_xmm}). The VLBA image reveals a dominant, marginally resolved source with a flux density of $4.46 \pm 0.04$\,mJy. 
This flux density is consistent with the  NVSS  measurement observed more than 15 years earlier but fainter than most contemporaneous data (Fig.~\ref{fig:sed}\textit{c}). At the moment, it is unclear whether the ``missing'' flux of the VLBA observations is due to variability or that we might be resolving out extended emission on a few pc scales. 
We obtained X-ray observations of \blazar\ with \textit{XMM-Newton} on 04 August 2022 and with \textit{Chandra} over six observations from 21 November 2022 to 21 December 2022 (see Methods and Fig.~\ref{fig:vlba_xmm}). When fit with a single model, the derived X-ray properties are a luminosity of $L_{2-10\ {\rm keV}}=3.2^{+0.8}_{-0.7}\ \times 10^{45}\ {\rm erg}\ {\rm s}^{-1}$, a photon index of   $\Gamma_X=1.47^{+0.19}_{-0.17}$,  and $\widetilde{\alpha}_{\rm OX}=-1.25\pm0.02$. Besides  satisfying the blazar classification\cite{ighina2019} $\Gamma_X<1.8$ and $\widetilde{\alpha}_{\rm OX}>-1.36$, we note that 
non-blazar quasars typically have $\Gamma_X\sim 1.9$ and that there is a strong indication that the $z>6$ quasars are softer\cite{zappacosta2023} with an average $\Gamma_X=2.4\pm 0.1$.

Based on a simple geometrical consideration\cite{volonteri2011,belladitta2020}, the existence of this blazar implies that there should be many more intrinsically similar jetted sources (e.g., similar black hole masses and redshifts) pointing elsewhere, including optically-obscured quasars. More specifically, the number of blazars with a viewing angle $\theta\leq 1/\Gamma$ (for $\Gamma \gg 1$), is related to the total jetted population, $N_{\rm jetted}$, as:
\begin{eqnarray}
    N(\theta\leq1/\Gamma) & = & N_{\rm jetted} \times \dfrac{2\Omega}{4\pi} \\
     & = & N_{\rm jetted} \int^{1/\Gamma}_0 \,\sin(\theta)\,d\theta\\
     & = & N_{\rm jetted} \left[1 - \cos\left(\frac{1}{\Gamma}\right)\right]
\end{eqnarray}
    
\noindent where $\Gamma$ is the bulk Lorentz factor of the emitting plasma, and $\Omega$ is the solid angle subtended by the jet. 
Using the first two terms of the Taylor series for $\cos(1/\Gamma)$, the total number of jetted sources given one blazar is $N_{\rm jetted}\simeq 2\Gamma^2$. Bulk Lorentz factors for blazars\cite{hovatta2009,saikia2016} range from 1 to 40. If $\Gamma$ is close to one, it could suggest a truly unique source. 
The distribution of bulk Lorentz factors for low-redshift blazars\cite{hovatta2009,hovatta2019} peaks between $\Gamma=5-15$, and at $z>4$, $\Gamma \approx 5$ provides a good match to observational constraints\cite{diana2022}.  
Assuming $\Gamma$ is in the range $4-15$, we expect about $30-450$ similar jetted sources to \blazar\ at $z\sim 7$. Thus far, there are two other spectroscopically confirmed jetted quasars\cite{banados2021, endsley2023MNRAS.520.4609E}  at a comparable redshift ($z\sim 6.8$) 
---implying many $z\sim 7$ radio sources are yet to be identified.

Quasars are observationally classified as radio-loud if their radio-loudness parameter\cite{kellermann1989,banados2021}---the ratio of rest-frame 5\,GHz (radio) to 4400\,\AA\ (optical) or 2500\,\AA\ (UV) ---is greater than 10. With the caveat that this source shows strong radio variability, we estimate its radio-loudness extrapolating the radio slopes measured from our two quasi-simultaneous 2021 and 2022 VLA observations to rest-frame 5\,GHz; i.e., with  $\alpha=0.85$ and $\alpha=0.21$, respectively (in the convention $f_{\nu} \propto \nu^{\alpha}$; see Fig.~\ref{fig:sed}\textit{a}).  This results in radio-loudness parameters  $R_{2500}=f_{5\,\mathrm{GHz}} / f_{2500\,\mathrm{A}}=165$ and $R_{4400}=f_{5\,\mathrm{GHz}} / f_{4400\,\mathrm{A}}=74$ in 2021 and $R_{2500}=331$ and $R_{4400}=148$ in 2022 (see Table~\ref{tab:properties}).

However, it is plausible that \blazar\ {has an intrinsically low-power, weak jet and that the observed radio loudness is only due to relativistic Doppler boosting. An enhancement of only 7.4 (14.8) would have made this quasar radio-quiet in 2021 (2022). The observed flux density is enhanced as   $f_{\nu,\mathrm{obs}} =\delta^{(3-\alpha)} \times  f_{\nu,\mathrm{int}}$, where $\alpha$ is the radio spectral index and $\delta$ is the relativistic Doppler factor $\delta = [\Gamma(1-\beta \cos(\theta))]^{-1}$ and $\beta = \sqrt{1-\left(1/\Gamma^2\right)}$. Fig.~\ref{fig:qlf_v2}\textit{a} is a Doppler enhancement diagram for \blazar, which shows a substantial chance that we see boosted flux (even reaching factors of $\sim 10^{3-4}$).

If the intrinsic luminosity of this radio jet is much lower than observed, the argument of the $2\Gamma^2$ similar sources still holds. 
However, the expected number of jetted sources approaches the total number of UV-bright quasars derived from the $z\sim 7$ quasar UV luminosity function (Fig.\ref{fig:qlf_v2}\textit{b}). The implication is that a large fraction of UV-bright quasars must have a relativistic radio jet even if they are classified as radio-quiet.    This aligns with the fact that some of the most massive quasars known at high redshift, although classified as radio-quiet, show evidence of relativistic radio jets\cite{sbarrato2021}. 
At low redshift, the origin of the radio emission for radio-quiet quasars is still debated, but there exist examples of relativistic jets powered by AGN that do not meet the observational criteria for classification as radio-loud\cite{macfarlane2021,wang2023MNRAS.518...39W}. 
We note that the expected contribution from star formation to radio emission is less than four orders of magnitude\cite{ceraj2018} of the observed radio luminosity of \blazar.
Confirming that a large fraction of the highest redshift quasars are jetted can have profound implications as relativistic jets {may affect the interstellar medium of their host galaxies\cite{venturi2021}, and enhance the growth of supermassive black holes\cite{jolley2008,ghisellini2013}. 

The radio-loud fraction of quasars is  constant\cite{ivezic2002,banados2015a} at $\sim10\%$ from $z<1$ to $z\sim 6$. Thus, if \blazar\ is intrinsically radio-loud ($R_{4400}>10$), it implies that there should be $\sim10$ times more radio-quiet quasars at similar redshifts. This strongly conflicts with the numbers expected from the quasar UV luminosity function\cite{matsuoka2023}, as shown in Fig.~\ref{fig:qlf_v2}\textit{b}. Since the number predicted from the detection of a blazar also includes obscured quasars, a way to reconcile the results is that most black hole growth must happen in an obscured phase.  Such obscured growth is predicted by some theoretical models\cite{johnson2022,satyavolu2023}, and it is observationally supported by the obscured $z=6.8$ quasar recently discovered in a small region of the sky\cite{endsley2023MNRAS.520.4609E} and obscured but intrinsically luminous AGN candidates uncovered by \textit{JWST}\cite{labbe2023,lambrides2024}.

The presence of \blazar\ implies that jet-enhanced\cite{jolley2008,connor2024} and/or obscured, sustained super-Eddington\cite{johnson2022} accretion can play a key role on supermassive black hole growth at early times. Both scenarios ease the tension regarding the existence of supermassive black holes in the early Universe, 
 leaving the collapse of dense star clusters and the remnants of Pop III stars as still viable options as initial ``seeds''.

\noindent \textbf{Acknowledgments:} 
We thank the multiple reviewers for their feedback, which improved the quality of the manuscript. 
We thank Rob Simcoe for conducting the Magellan/FIRE observation in September 2021.  
We thank Melanie Habouzit and Bhargav Vaidya for their insightful discussions and feedback on this paper.  
C.M.\ acknowledges support from Fondecyt Iniciacion grant 11240336 and ANID BASAL project FB210003.
Y.K.\ thanks the support of the German Space Agency (DLR) through the program LEGACY 50OR2303. 
J.-T.S.\ is supported by the Deutsche Forschungsgemeinschaft (DFG, German Research Foundation) - Project number 518006966. 
A.P.\ acknowledges support from Fondazione Cariplo grant no.\ 2020-0902. 
We acknowledge using the DELS, NVSS, and VLASS surveys to identify the quasar.  
This work is based on data collected with the Very Large Telescope (program 106.20WQ.001), the New Technology Telescope (programs 106.20WJ.001, 105.203B.001, 110.23RN.001),  the Large Binocular Telescope (program MPIA-2021B-003), the Keck Telescope, the Magellan Baade Telescope, the Very Large Array (program VLA/21B-087), the Very Large Baseline Array (program VLBA/21B-190), \textit{XMM-Newton} (ObsID 08916), and \textit{Chandra} (Seq Num 704639).
To construct the SED of $z\sim1$ blazars shown in Fig.~\ref{fig:sed}, we use 
SED builder (\url{https://tools.ssdc.asi.it/SED/}); thus, part of this work is based on archival data, software or online services provided by the Space Science Data Center - ASI.
We are grateful for the support provided by the staff of these observatories. 
The authors wish to recognize and acknowledge the very significant cultural role and reverence that the summit of Maunakea has always had within the indigenous Hawaiian community.  We are most fortunate to have the opportunity to conduct observations from this mountain. The National Radio Astronomy Observatory is a facility of the National Science Foundation operated under cooperative agreement by Associated Universities, Inc.

\medskip
\noindent \textbf{Author contributions:} 
E.B.\ led the team and wrote the main manuscript with E.M.\ and T.C. E.B.\ selected the quasars and with C.M.\ and B.P.V\ obtained and analyzed the discovery spectrum. E.B.\ led the FORS2, LUCI, VLA, and XMM-Newton proposals and analyzed the spectra, the NTT and VLA data. E.M. led the VLBA proposal, reduced the VLA and VLBA data, and performed the VLBA analysis. T.C.\ led the Chandra proposal and reduced and analyzed the Chandra and XMM-Newton data.    R.D\ led the NTT proposals and reduced the data. A-C.E.\ led the FIRE proposal and reduced the data. A.J.B.\ and J.F.H.\ carried out the NIRES observations and F.Wang reduced the data. 
 Z-L.X.\ reduced the FORS2 and LUCI spectra. S.B.\ compiled all the archival radio data. All co-authors discussed the results and provided input to the data analysis and the content of the paper and telescope proposals.   
 Correspondence and requests for materials should be addressed banados@mpia.de.

\medskip
\noindent \textbf{Competing interest:} 
The authors declare no competing financial and non-financial interests.

\newpage

\clearpage
\section*{Methods}

\subsection*{VLASS flux densities}
\label{sec:vlass}
For the identification and radio characterization of \blazar, we used peak flux densities from the first data products of the VLASS survey\cite{lacy2020}: the VLASS Quick Look (QL) images. The QL images have some known systematic offsets of their flux densities (\url{https://science.nrao.edu/vlass/data-access/vlass-epoch-1-quick-look-users-guide}). In practice, the peak flux densities of QL VLASS1.1 data are low by $\approx 15\%$, while for VLASS2.1 and VLASS3.1 by $\approx 8\%$. We have taken these factors into account in the reported values in Table~\ref{tab:obs}. 

During the revision of this paper, the VLASS2.1 Single Epoch (SE) continuum image of \blazar\ was released. The SE images are intended to be the VLASS reference data products and supersede the QL images (\url{https://science.nrao.edu/vlass/vlass-se-continuum-users-guide}.) 
Therefore, we report the SE VLASS2.1 measurement in Table~\ref{tab:obs}. The difference between the SE and QL flux densities is $0.49 \pm 0.24\,$mJy.

\subsection*{Data and data reduction}
\subsubsection*{NIR imaging}

We obtained two epochs of \Jntt, \Hntt, and \Kntt\ imaging 
with SOFI \cite{moorwood1998} at the NTT telescope in La Silla on 20 November 2020 and 04 January 2023. 
An additional epoch of \Jntt\ only was obtained on 26 July 2021. 
The data were bias-subtracted, flat-fielded, sky-subtracted, and stacked. The photometric zero points were calibrated using stars from the 2MASS\cite{skrutskie2006} survey. 
The NTT/SOFI photometry is reported in Table~\ref{tab:obs}; the photometry from 2020 and 2023 are consistent within the uncertainties, while the \Jntt\ photometry from 2021 is slightly fainter by only $0.2$\,mag, but consistent within $2\sigma$.

\subsubsection*{Spectroscopy}

We obtained a 30 min spectrum on 07 November 2020 with the FOcal Reducer/low dispersion Spectrograph~2 (FORS2\cite{appenzeller1992}) at the Very Large Telescope (VLT). 
The FORS2 spectrum detected a strong Ly$\alpha$ line and a sharp Lyman break, confirming the quasar nature of \blazar. On 26 December 2020, we obtained a 1.2\,h Keck/NIRES\cite{wilson2004} spectrum, which revealed strong \civ\ and \mgii\ broad emission lines.  
We observed \blazar\ on 23 September 2021 and 15 October 2021 for a total of 6\,h with Folded-port Infrared Echellete (FIRE\cite{simcoe2013}) spectrograph in echelle mode at the Magellan Baade telescope at Las Campanas Observatory. 
On 23 October 2021 and 12 November 2021, we obtained fraternal mode spectroscopy with the LUCI\cite{buschkamp2012} spectrograph at the Large Binocular Telescope (LBT), where $zJ$ and $HK$ observations are carried out simultaneously.  The November observations were taken under good weather conditions, while bad weather in October severely affected the data. We discarded all individual frames with a S/N$<$0.1, which eliminated all the October $HK$ data.  The total effective exposure time for the $zJ$ and $HK$ spectra are 5 and 3.5\,h, respectively.

All the spectra were reduced with the Python Spectroscopic Data Reduction Pipeline (PypeIt\cite{prochaska2020}). 
The LBT/LUCI and Magellan/FIRE data were absolute flux calibrated to match the $Ks_{\rm SOFI}=20.30$ photometry. The VLT/FORS2 and Keck/NIRES spectra were scaled to match the flux-calibrated Magellan/FIRE spectral continuum.
Fig.~\ref{fig:discovery} shows the individual and combined spectra, revealing strong, broad emission lines typical of blazars that are not BL Lac sources\cite{sbarrato2013,ghisellini2015c}. The continuum slope and profile of the broad emission lines are consistent across all the different spectra. 
Stable broad line regions are also seen in low redshift blazars with multi-epoch spectroscopic campaigns\cite{carnerero2015,raiteri2020}.

\subsubsection*{VLA}
To characterize the radio spectral energy distribution of the quasar \blazar, we carried out observations with the Karl G. Jansky Very Large Array (VLA) of the NRAO in its B-configuration (maximum baseline = 11.1\,km) on 25 September 2021. The frequency range of the VLA observations spanned 1 to 12\,GHz using the L- (1-2 GHz), S- (2–4 GHz), C- (4–8 GHz), and X- (8-12) band receivers. The total on-source time was $\sim 20$\,min at L- and S-bands, and $\sim 22$\,min at C- and X-bands. The source J0542+4951 (3C\,147) was observed as the flux density scale calibrator and bandpass calibrator, while the source J0423$-$0120 was observed to calibrate the complex gains. The L- and S-band observations used the 8-bit samplers of the VLA, while the C- and X-bands used the 3-bit samplers. The Wideband Interferometric Digital ARchitecture (WIDAR)
correlator was configured to use the default NRAO wide-band setups that deliver 16 $\times$ 64 MHz sub-bands at L-band, 16 $\times$ 128 MHz sub-bands at S-band, and 32 $\times$ 128 MHz sub-bands at both C- and X-bands. Data editing, calibration, deconvolution, and imaging were carried out using the Common Astronomy Software Applications (CASA\cite{mcmullin2007}) package. Self-calibration in phase only for L- and C- bands and in both phase and amplitude in S- and X-bands was also performed on the target source to further improve the dynamic range and the image fidelity. Self-calibration at L-, S-, and C-bands was carried out in CASA. The X-band data were self-calibrated in phase using CASA, while the amplitude self-calibration required the use of the Astronomical Image Processing System (AIPS\cite{greisen2003}) 
to properly constrain the amplitude gain values.

Another epoch of VLA L-, S-, C- and X-band observations on the target source were carried out in the D-configuration (maximum baseline = 1\,km) on 6 August 2022. The setup and the data reduction were similar to those used for the B-configuration observations. The flagging of Radio Frequency Interference resulted in slightly different effective central frequencies between the 2021 and 2022 data (see Table~\ref{tab:obs}).

The final images were all made in CASA using a weighing scheme intermediate between uniform and natural (robust=0.4 in CASA \texttt{tclean}), and the multi-terms multi-frequency synthesis (\texttt{MTMFS}) algorithm with two terms (nterms=2 in CASA \texttt{tclean}). The latter allows us to model both the total intensity and the spectral index variations in the sky. These final images were centered at 1.5\,GHz (L-band data), 3\,GHz (S-band data), 5 and 7\,GHz (C-band data) 8.9 and 10.7\,GHz (X-band data). The synthesized beam sizes at FWHM are reported in Table~\ref{tab:obs}. 

A 2-dimensional Gaussian fitting on the source at all these frequencies and epochs reveals a pure point source. 
The measured flux densities are listed in Table~\ref{tab:obs}.

\subsubsection*{VLBA}
\blazar\ was observed with the Very Long Baseline Array (VLBA) of the NRAO at
1.5 GHz (L-band) on 17 Nov 2021. Eight 32 MHz subband pairs were recorded at each station using the ROACH Digital Backend and the polyphase filterbank (PFB) digital signal-processing algorithm, both with right- and left-hand circular polarizations, and sampled at two bits. The
total bandwidth was 256 MHz, centered at 1.54 GHz. The total observing time at L-band was 6\,h, with 4.1\,h on target.

The VLBA observations utilized nodding-style phase referencing using the calibrator source J0408$-$0122, which is $0.54^{\circ}$ from the target. The phase referencing cycle time was 5 minutes: 4 minutes on the target and 1 minute on the calibrator. The uncertainty in the calibrator’s position is 0.13 milliarcsec (mas) in right ascension and 0.27 mas in declination. 
As employed in these observations, phase referencing would preserve absolute astrometric positions\cite{fomalont1999} to better than $0.01^{\prime\prime}$. 
The observations also included the calibrator source 3C\,84, which was used as a fringe finder and bandpass calibrator. Amplitude calibration was performed using measurements of the antenna gain and the system temperature of each station. The data were correlated with the VLBA DiFX correlator\cite{deller2011} 
in Socorro, NM, with 1\,s correlator
integration time. 
Data reduction and analysis were performed in AIPS
following standard VLBI data reduction procedures. The phase reference source was self-calibrated and the solutions were applied on the target field.
Deconvolution and imaging of the target were performed using natural weighing (Robust=5 in AIPS task \texttt{IMAGR}).

Fig.~\ref{fig:vlba_xmm} shows the VLBA 1.5\,GHz (12\,GHz rest frame) image of J0410$-$0139 at an angular resolution of $14.3 \times 5.9$ mas ($ 75 \times 37$ pc) with a position angle of P.A. = $1^{\circ}$. The rms noise in the image is 20\,$\mu$Jy\,beam$^{-1}$. 
This image shows a dominant, marginally resolved source with a flux density of $4.46 \pm 0.04$\,mJy, and a size 
$< 3$\,mas ($<16$\,pc). The derived intrinsic brightness temperature lower limit is $2 \times 10^9$\,K.

\subsubsection*{X-ray Observations}
We observed \blazar\ with \textit{XMM-Newton} on 04 August 2022 
for 55~ks using the European Photon Imaging Camera (EPIC\cite{struder2001,turner2001}). Observations were reduced with the \textit{XMM} Science Analysis Software (SAS) v19.0.0, and, after filtering for periods of high background, we had effective exposure times of 28.2 ks (MOS1), 28.6 ks (MOS2), and 13.6 ks (pn). A combined image of these observations is shown in Figure \ref{fig:vlba_xmm}; there are no other notable X-ray sources in the immediate vicinity of the quasar. We used a circular extraction region of radius 20$^{\prime\prime}$ centered on the quasar for spectral analysis. The resultant spectra were binned to a minimum of 5 counts per bin. 

Additionally, we observed \blazar\ with \textit{Chandra} in six observations totaling 132~ks, starting 25 November 2022 and finishing 21 December 2022. \blazar\ was positioned on the ACIS-S3 chip, and data were collected with the Very Faint telemetry format. \textit{Chandra} observations were reduced and processed with Chandra Interactive Analysis of Observations (CIAO) v4.15\cite{2006SPIE.6270E..1VF}, and we used a circle of radius $3^{\prime\prime}$ for spectral extraction. As before, we binned the combined spectrum to a minimum of 5 counts per bin. Variability between the two observation sets is below the statistical uncertainties.

Spectral analysis was performed using XSPEC\cite{ArnaudXSPEC} to minimize the C-stat\cite{Cash1979,Wachter1979}. We modeled the X-ray emission as a power law with a Galactic absorption component \cite{HI4PI2016} of $6.99\times10^{20}\ {\rm cm}^{-2}$. When jointly fitting both observations with the same model, the best fit ($C$/DoF = 218.54/205) was found for a power-law index of $\Gamma_X=1.47^{+0.19}_{-0.17}$ and 1 keV flux density $S_{1\ {\rm keV}}=1.90^{+0.31}_{-0.30}\ {\rm nJy}$, corresponding to an intrinsic X-ray luminosity of $L_{2-10\ {\rm keV}}=3.2^{+0.8}_{-0.7}\ \times 10^{45}\ {\rm erg}\ {\rm s}^{-1}$ and a modeled flux of $f_{0.5-7.0\ {\rm keV}}=1.71^{+0.22}_{-0.19}\times 10^{-14}\ {\rm erg}\ {\rm s}^{-1}\ {\rm cm}^{-2}$. From this fit and the rest-frame UV luminosity reported above, we find values 
of $\alpha_{\rm OX}=-1.46\pm0.07$ and $\widetilde{\alpha}_{\rm OX}=-1.25\pm0.02$ (see Table~\ref{tab:properties} for their definitions).  We note that \blazar's $\Gamma_X$ and $\widetilde{\alpha}$ values are consistent with what is expected for blazars\cite{ighina2019}, and in contrast to most of the X-ray properties of other $z>6$ quasars\cite{zappacosta2023}.

\medskip
\medskip
\medskip

\noindent \textbf{Data availability:} The data used in this study can be accessed from the observatories public archives or the surveys websites. The datasets generated and analyzed during this study are available from the corresponding author upon reasonable request. \\
\medskip
\noindent \textbf{Inclusion \& Ethics :} N/A. \\
\medskip
\noindent \textbf{Code availability :} N/A. \\

\clearpage



\section*{Tables}
\renewcommand{\thetable}{\arabic{table}} 
\setcounter{table}{0}

\begin{table}[h!]
\small
\centering

\caption{\textbf{Optical and near-infrared photometry  of \blazar}}
\begin{tabular}{lll}
  \hline 
  Survey/Inst.  & AB magnitudes & Date \\
  \hline
  DELS DR10  & $g_{\rm DE, 3\sigma}>25.66$&            \\
  DELS DR10  & $r_{\rm DE, 3\sigma}>25.02$&            \\
  DELS DR10  & $i_{\rm DE, 3\sigma}>24.45$&            \\
  DELS DR10  & $Z_{\rm DE}=22.14\pm 0.06$&    \\
  NTT/SOFI  & $J_{\rm SOFI}=20.75\pm 0.07$& 2020-11-20 \\
  NTT/SOFI  & $J_{\rm SOFI}=20.97\pm 0.12$& 2021-07-26 \\
  NTT/SOFI  & $J_{\rm SOFI}=20.77\pm 0.07$& 2023-01-04 \\
  NTT/SOFI  & $H_{\rm SOFI}=20.85\pm 0.13$& 2020-11-20 \\
  NTT/SOFI  & $H_{\rm SOFI}=20.88\pm 0.07$& 2023-01-04 \\
  NTT/SOFI  & $Ks_{\rm SOFI}=20.30\pm 0.08$& 2020-11-20 \\
  NTT/SOFI  & $Ks_{\rm SOFI}=20.38\pm 0.10$& 2023-01-04 \\
  DELS DR10  & $W1=20.46\pm0.08$&            \\
  DELS DR10  & $W2=20.16\pm0.15$&            \\
  \hline
  \label{tab:obs}
  \end{tabular}
\\
\smallskip
\footnotesize{ \textbf{Notes.}  For the selection of \blazar, we used DELS DR8, but here we report the most recent photometry from DELS~DR10.}
\end{table}

\begin{table}[h!]
\small
\centering

\caption{\textbf{Radio observations of \blazar}}
\begin{tabular}{lllll}
  \hline 
Frequency (GHz) & Flux density (mJy) & Date & Image beam (Pos.\ Angle) & Reference \\
  \hline
0.856     & $5.09 \pm 0.30$ & 2021-12-18 & $23\farcs1 \times  11\farcs8$ ($-61^\circ$) & FLASH\cite{allison2022}      \\
0.888     & $5.64 \pm 0.90$ & 2019-04-26 & $25\farcs0 \times  25\farcs0$ ($0^\circ$) & 
 RACS\cite{hale2021}  \\
0.888     & $5.42\pm 0.97$         & 2020-06-20 & $15\farcs1 \times  12\farcs4$ ($85^\circ$) &  VAST\cite{murphy2021}  \\
0.888     & $5.10 \pm 0.8$  & 2021-07-21 & $13\farcs6 \times  11\farcs9$ ($79^\circ$) &  VAST\cite{murphy2021}  \\
1.37     & $7.89 \pm 1.62$  & 2021-07-28 & $9\farcs4 \times  7\farcs6$ ($88^\circ$) &  VAST\cite{murphy2021}  \\
1.37     & $11.99 \pm 1.84$  & 2021-09-20 & $12\farcs9 \times  7\farcs1$ ($67^\circ$) &  VAST\cite{murphy2021}  \\
1.37     & $5.83 \pm 1.5$  & 2021-11-18 & $10\farcs6 \times  7\farcs8$ ($81^\circ$) &  VAST\cite{murphy2021}  \\
1.40      & $4.20 \pm 0.50$ & 1995       & $45\farcs0 \times  45\farcs0$ ($0^\circ$) &  NVSS\cite{condon1998} \\
1.5       & $8.12 \pm 0.02$ & 2021-09-25 & $4\farcs26 \times  3\farcs45$ ($-18^\circ$) &  VLA/21B-087 \\
1.5       & $4.46 \pm 0.04$ & 2021-11-17 & $0\farcs0143 \times  0\farcs0059$ ($1^\circ$) &  VLBA/21B-190 \\
1.5       & $9.21 \pm 0.17$ & 2022-08-06 &  $40\farcs5 \times  34\farcs5$ ($3^\circ$) & VLA/21B-087 \\
3.0       & $7.60 \pm 0.20$ & 2017-10-01 & $2\farcs8 \times  2\farcs1$ ($16^\circ$) & VLASS1.1\cite{lacy2020} \\
3.0       & $10.29 \pm 0.14$ & 2020-07-19 & $2\farcs8 \times  2\farcs1$ ($-1^\circ$)& VLASS2.1\cite{lacy2020} \\
3.0       & $14.48 \pm 0.01$ & 2021-09-25 & $2\farcs16 \times  1\farcs83$ ($-9^\circ$)& VLA/21B-087 \\
3.0       & $10.63 \pm 0.02$ & 2022-08-06 &  $23\farcs13 \times  19\farcs53$ ($2^\circ$) & VLA/21B-087 \\
3.0       & $10.10 \pm 0.15$ & 2023-03-08 & $2\farcs8 \times  2\farcs1$ ($1^\circ$)& VLASS3.1\cite{lacy2020} \\
5.0       & $15.70 \pm 0.01$ & 2021-09-25 & $1\farcs37  \times  1\farcs02$ ($-8^\circ$) & VLA/21B-087 \\
5.0       & $11.05 \pm 0.01$ & 2022-08-06 & $16\farcs0  \times  11\farcs6$ ($-0^\circ$) & VLA/21B-087 \\
6.0       & $14.88 \pm 0.01$ & 2021-09-25 &  $1\farcs 23\times 0\farcs 89$ ($-2^\circ$) & VLA/21B-087 \\
6.0       & $10.66 \pm 0.01$ & 2022-08-06 &  $13\farcs0\times 9\farcs 0$ ($7^\circ$) & VLA/21B-087 \\
7.0       & $14.04 \pm 0.01$ & 2021-09-25 &  $1\farcs 13  \times  0\farcs 79$ ($2^\circ$) &VLA/21B-087 \\
7.0       & $10.28 \pm 0.01$ & 2022-08-06 &  $11\farcs5  \times  8\farcs 0$ ($10^\circ$) &VLA/21B-087 \\
8.74       & $9.45 \pm 0.01$ & 2022-08-06 &  $9\farcs 0  \times  0\farcs58$ ($-4^\circ$) &VLA/21B-087 \\
8.87       & $12.79 \pm 0.01$ & 2021-09-25 &  $0\farcs 77  \times  6\farcs9$ ($3^\circ$) &VLA/21B-087 \\
10.0       & $12.01 \pm 0.005$ & 2021-09-25 &  $0\farcs69\times 0\farcs 52$ ($-4^\circ$) & VLA/21B-087 \\
10.0       & $8.87 \pm 0.06$ & 2022-08-06 &  $0\farcs77 \times 0\farcs58$ ($-4^\circ$) & VLA/21B-087 \\
10.19      & $8.79 \pm 0.01$ & 2022-08-06 &  $7\farcs 8  \times  5\farcs9$ ($11^\circ$) &VLA/21B-087 \\
10.7      & $11.54 \pm 0.01$ & 2021-09-25 &  $0\farcs 64  \times  0\farcs49$ ($-4^\circ$) &VLA/21B-087 \\
  \hline
  \label{tab:radio}
  \end{tabular}

\smallskip
\footnotesize{ \textbf{Notes.}  The peak flux densities of the QL VLASS1.1 are corrected by $15\%$, while for VLASS2.1 and VLASS3.1 by $ 8\%$ (see Methods).}

\end{table}

\begin{table}[h!]
\small
\centering
\caption{\textbf{Derived properties of \blazar.}}
\begin{tabular}{lll}
\hline 
Quantity  & Value & Units \\
\hline
\vspace{2pt}
$z_{\rm \mgii}$     & $6.995\pm 0.001$         &          \\
$z_{\rm \cii}$      & $6.9964\pm 0.0005$         &          \\
$m_{1450}$      & $21.33\pm 0.08$         &  AB mag         \\
$M_{1450}$      & $-25.60\pm 0.08$         &  AB mag         \\
$L_{2500}^a$      & $(2.0\pm 0.1)\times 10^{46}$         &  erg\,s$^{-1}$        \\
$L_{3000}^a$      & $(2.1\pm 0.1)\times 10^{46}$         &  erg\,s$^{-1}$        \\
$L_{4400}^a$      & $(2.5\pm 0.2)\times 10^{46}$         &  erg\,s$^{-1}$        \\
$L_{5\mathrm{\,GHz}}(2021)^b$      & $(13.590\pm 0.003)\times 10^{42}$         &  erg\,s$^{-1}$        \\
$R_{2500}(2021)^b$      & $165\pm 11$         &        \\
$R_{4400}(2021)^b$      & $74\pm 5$         &        \\
$L_{5\mathrm{\,GHz}}(2022)^b$      & $(27.210\pm 0.006)\times 10^{42}$         &  erg\,s$^{-1}$        \\
$R_{2500}(2022)^b$      & $331\pm 22$         &        \\
$R_{4400}(2022)^b$      & $148\pm 9$         &        \\
FWHM$_{\rm \mgii}$      & $2562^{+90}_{-80}$        & \kms      \\
$M_{\rm BH}$            & $6.9^{+0.5}_{-0.4} \times 10^8$ & $M_\odot$   \\
$L_{\rm bol}$      & $(1.096\pm 0.005)\times 10^{47}$         &  erg\,s$^{-1}$        \\
$L_{\rm bol}/L_{\rm Edd}$      & $1.22\pm 0.08$        &      \\
 $L_{2-10\ {\rm keV}}$ &  $3.2^{+0.8}_{-0.7}\ \times 10^{45}$ & erg\,s$^{-1}$\\
$\Gamma_X$ & $1.47^{+0.19}_{-0.17}$ &   \\
$\alpha_{\rm OX}$$^c$ & $-1.46 \pm 0.07 $ &   \\
$\widetilde{\alpha}_{\rm OX}$$^c$ & $-1.25 \pm 0.02$&   \\
\hline
\label{tab:properties}
\end{tabular}
\\

\footnotesize{
$^a$ The UV/optical luminosities are measured from the best-fit power-law model to the combined spectrum. The uncertainties are dominated by the $Ks_{\rm SOFI}=20.30\pm 0.08$ magnitude used to absolute flux-calibrate the spectrum. 
\\
$^b$ The rest-frame 5\,GHz luminosity and radio loudness are reported using the radio measurements from 2021-09-25 and 2022-08-06; see Table~\ref{tab:obs} and Fig.~\ref{fig:sed}a.\\
$^c$ $\alpha_{\rm OX} \equiv \log(L_{\nu_X} / L_{\nu_{\rm UV}}) / \log(\nu_X / \nu_{\rm UV})$, where $L_\nu$ is the monochromatic luminosity. In this work, and in keeping with standard practices, we evaluate $\nu_{\rm UV}$ at 2500 \AA\ and $\nu_X$ at 2 keV ($\alpha_{\rm OX}$) and 10 keV ($\widetilde{\alpha}_{\rm OX}$).
} 
\end{table}
\clearpage
\newpage
\section*{Figure captions}

\renewcommand{\thefigure}{\arabic{figure}}
\setcounter{figure}{0}

\begin{figure}[h!]
\centering
\includegraphics[]{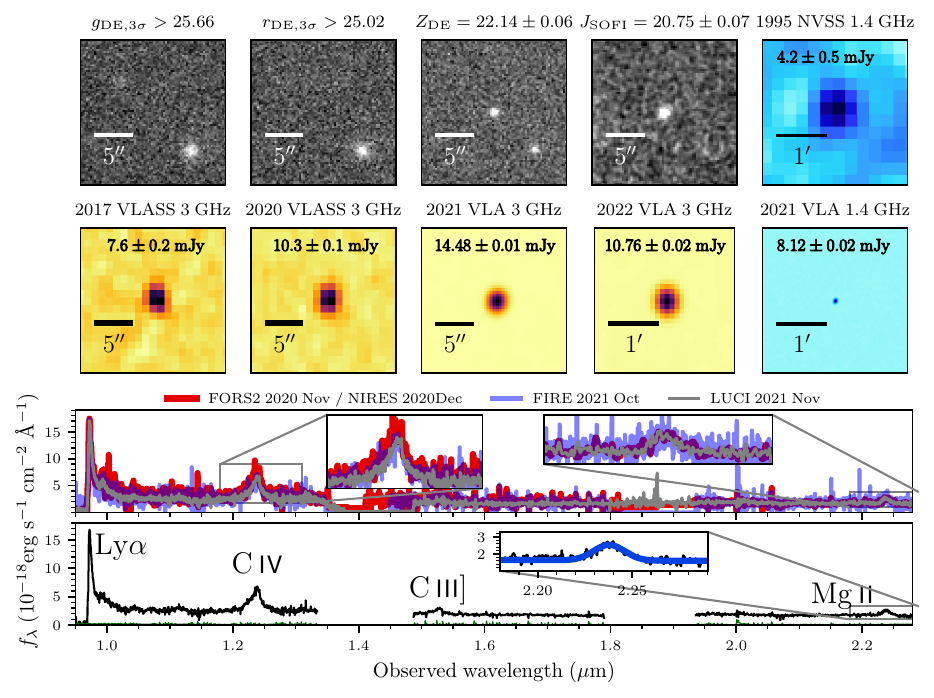}
\caption{\small
\textbf{Photometry and spectra of the quasar \blazar\ at $z=7.0$.} 
The grayscale postage stamps are optical and near-infrared images; radio images use different colormaps (3\,GHz in orange, 1.4\,GHz in light blue).  The 1995 NVSS 1.4\ and the 2022 VLA 3\,GHz images, taken in the most compact VLA configuration, have the lowest resolution. The 2021 VLA 1.4\,GHz image, shown in the same size for comparison, confirms a single radio source in the field with strong evidence of variability. 
 The second-to-bottom panel compares spectra from VLT/FORS2, Keck/NIRES, Magellan/FIRE, and LBT/LUCI, showing consistent rest-frame UV with no variability in the broad line region or accretion disk. 
 The bottom panel shows the combined spectrum (and the $1\sigma$ error in green) with a Gaussian fit to the \mgii\ line, used to estimate the black hole properties (Table~\ref{tab:properties}).
}
\label{fig:discovery}
\end{figure}

\begin{figure}[h!]
\centering
\includegraphics{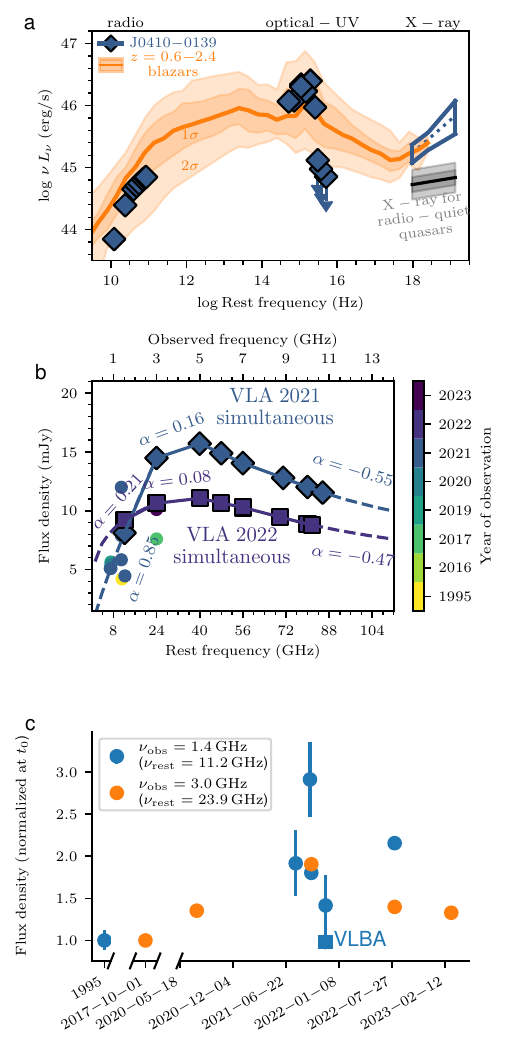}
\caption{\small
\textbf{SED and radio lightcurve of \blazar}. \textbf{(a)} Rest-frame SED from radio to X-rays, showing only the VLA 2021 simultaneous observations. Strong absorption in the UV (shown as $3\sigma$ upper limits) is due to absorption from the neutral intergalactic medium at $z\sim 7$. The X-ray uncertainties consider $2\sigma$ variation in the X-ray photon index and flux densities. The shaded orange regions span the 1 and $2\sigma$ SED distribution for low-redshift FSRQ blazars\cite{massaro2015} with similar X-ray luminosities, with the orange curve marking the median SED. The gray shading represents the 1 and 2$\sigma$ expected X-ray luminosity for a radio-quiet quasar with the same $L_{2500}$\cite{lusso2016}. 
\textbf{(b)} Radio SED with data points color-coded by year (Table~\ref{tab:obs}). The quasar shows strong variability across radio frequencies. Diamonds (squares) represent quasi-simultaneous VLA 2021 (2022) observations, within 15 minutes in the rest frame. Dashed lines are power-law extrapolations from these measurements. \textbf{(c)} Observed 1.4 and 3.0\,GHz (11.2 and 23.9\,GHz rest-frame) radio lightcurve. Data points are normalized to the initial flux density for each frequency, with $1\sigma$ error bars shown. The 1.4\,GHz values are corrected using spectral index values of $\alpha=0.85$ (pre-2022) and $\alpha=0.21$ (post-2022). 
\label{fig:sed}
}
\end{figure}

\begin{figure}[ht!]
\centering
\includegraphics[]{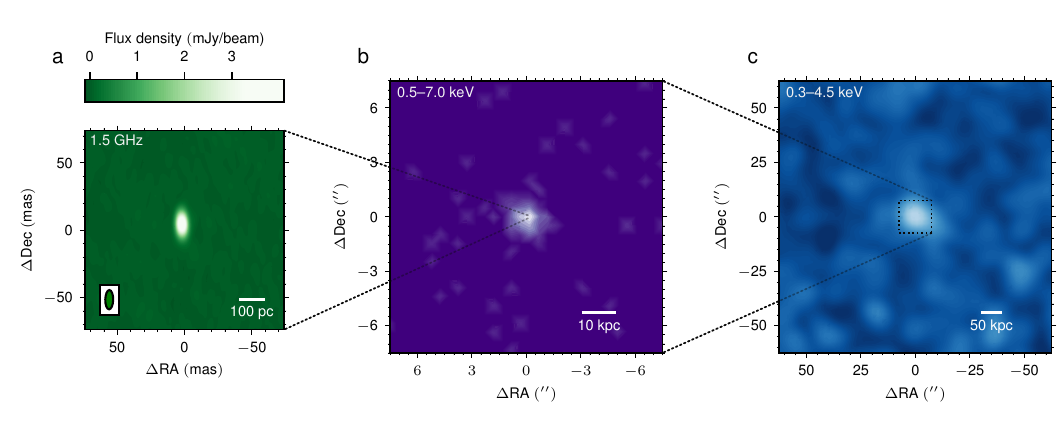}
\caption{\small 
\textbf{High-resolution 1.5\,GHz radio and X-ray observations of \blazar.} \textbf{(a)} VLBA 1.5\,GHz image with a beam size of $14.3 \times 5.9$ mas ($ 75 \times 37$ pc). 
The source is marginally resolved with a size $< 3$\,mas ($<16$\,pc). \textbf{(b)} \textit{Chandra}/ACIS 0.5--7.0\,keV image. \textbf{(c)} \textit{XMM}/EPIC 0.3--4.5\,keV image. 
\blazar\ is unresolved in the \textit{XMM} and \textit{Chandra} images and is the strongest X-ray source in the field.
VLBA and X-ray images reveal a single compact dominant source, supporting the blazar interpretation. 
}\label{fig:vlba_xmm}
\end{figure}

\begin{figure}[h!]
\centering
\includegraphics{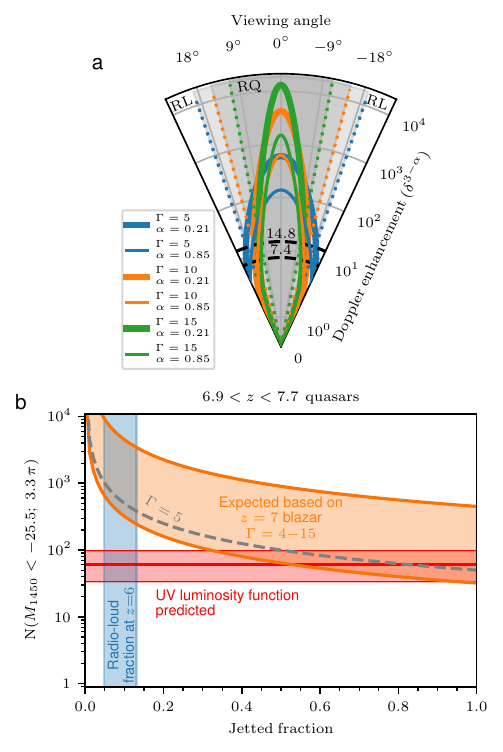}
\caption{\small
\textbf{Doppler enhancement and tension with the expected number of quasars at}\ $\mathbf{z\sim7}$. \textbf{(a)} 
Doppler enhancement diagram for different Bulk Lorentz Factor ($\Gamma$) and radio spectral indices ($\alpha=0.85$ for 2021 and $\alpha=0.21$ for 2022). \blazar\ is classified as radio-loud with $R_{4400}=74\, (148)$, but its jet may be intrinsically weak depending on relativistic Doppler enhancement. The dashed line indicates a Doppler enhancement of 7.4 (14.8), critical for reclassifying the quasar as radio-quiet in 2021 (2022). The shaded region represents viewing angles where \blazar\ would be intrinsically weak for the respective $\Gamma$. For larger angles, the 
jet power would be substantial, maintaining its classification as radio-loud. 
\textbf{(b)} 
Expected number of $6.9<z<7.7$ quasars brighter than $M_{1450}=-25.5$ over 3.3$\pi$ of the sky (covered by NVSS and VLASS), as a function of the fraction of quasars with relativistic jets. The horizontal red line represents the expected number based on the $z \sim 7$ quasar UV luminosity function, while the shaded region indicates the $1\sigma$ uncertainty\cite{matsuoka2023}. 
The existence of this blazar suggests $\sim2\Gamma^2$ quasars with similar properties but misaligned jets. The orange region shows the expected number of comparable quasars assuming typical low-redshift blazar values\cite{saikia2016} of $\Gamma=4-15$ as a function of the jetted fraction ($f_{\rm jetted}$): $N_{\rm total} = N_{\rm jetted}/f_{\rm jetted} = 2\Gamma^2/f_{\rm jetted}$. The dashed line represents an average $\Gamma=5$, consistent with observations\cite{diana2022} at $z>4$.  This estimate is in tension with the UV luminosity function. 
The shaded blue region shows the measured radio-loud fraction\cite{banados2015a} at $z=6$. 
If \blazar's jet is weak but appears radio-loud due to beaming, the tension decreases if the jetted fraction of UV-bright quasar exceeds 70\%. Conversely, if \blazar's jet is powerful and 
the radio-loud fraction remains $10\%$ at $z\sim 7$, the total number of quasars would exceed expectations by at least an order of magnitude.
}
\label{fig:qlf_v2}
\end{figure}

\end{document}